# Self-Viscophoresis: Autonomous Motion by Biasing Thermal Fluctuations via Self-Generated Viscosity Asymmetry

Bokusui Nakayama,[1] Yusuke Takagi,[2] Ryoya Hirose,[1] Masatoshi Ichikawa,[3]
Marie Tani,[1] Ibuki Kawamata,[1] Eiji Yamamoto,[3] and Akira Kakugo[1,*]

[1]Graduate School of Science, Kyoto University, Kitashirakawa, Sakyo, Kyoto, 606-8502, Japan
[2]Department of System Design Engineering, Keio University, 3-14-1 Hiyoshi, Kohoku, Yokohama, Kanagawa, 223-8522, Japan
[3]Graduate School of Integrated Sciences for Life, Hiroshima University, 1-3-2 Kagamiyama, Higashi-Hiroshima City, Hiroshima, 739-8511, Japan

(Dated: February 16, 2026)

Microscale transport often relies on ubiquitous yet intrinsically random thermal fluctuations. Understanding how such fluctuations can be biased into directed motion has long been a central theme of nonequilibrium physics. Here, we introduce self-viscophoresis, a mechanism of autonomous motion based on the rectification of thermal fluctuations in a self-generated nonequilibrium viscosity field. Asymmetric colloidal particles dispersed in a thermoresponsive polymer solution induce local heating under uniform illumination, producing a spatially asymmetric viscosity profile around the particle and resulting in persistent directed motion. To elucidate the physical origin of this behavior, we develop a minimal Langevin model coupling isotropic thermal fluctuations to a dynamically updating temperature–viscosity field. The model shows that viscosity asymmetry anisotropically damps stochastic dynamics, effectively biasing thermal fluctuations into a net drift. It thus reproduces the observed directed motion without invoking deterministic propulsion terms associated with effective potentials or environmental fluid flows. Our results distinguish self-viscophoresis from conventional self-propulsion mechanisms and establish it as a general framework enabling reversible control of both the direction and dimensionality of motion.

## INTRODUCTION

While thermal fluctuations typically randomize microscale motion, they can be rectified into directed transport in nonequilibrium systems[1,2]. A widely studied mechanism of fluctuation rectification is the Brownian ratchet[3–7], where a spatially asymmetric and temporally varying potential acts as a direction-dependent filter that converts isotropic thermal fluctuations into directional transport. In these systems, although directed motion effectively arises from thermal fluctuations under nonequilibrium operation, the required asymmetry is externally imposed through a prepatterned energy landscape[8–10]. Crucially, the transported particles here cannot generate or modulate this asymmetry themselves and thus remain passive cargoes in their own motion.

In contrast, some biological systems employ self-regulated strategies; they actively modify their rheological environments to create spatial asymmetry in mobility or diffusivity. Notably, certain bacteria can locally and nonuniformly modulate their surrounding viscoelastic medium. This rheological asymmetry selectively passes thermal fluctuations in favorable directions, thereby enabling directed motion even without self-propelling structures such as flagella[11–13]. In such systems, the direction of motion is governed by the organism's internal orientation or configuration, enabling autonomous behavior in heterogeneous and dynamically changing environments.

Realizing autonomous motion based on the rectification of thermal fluctuations in controllable systems is essential for understanding how nonequilibrium asymmetry biases stochastic motion and for establishing general design principles for active matter systems. In the active matter community, various self-propelled particles and propulsion mechanisms in artificial systems have been proposed. However, most of them rely on external forces[14,15] or momentum exchange with the surrounding solvent[16–22], in which stochastic thermal fluctuations are generally treated as mere background noise rather than as an integral part of the transport mechanism. Consequently, the fundamental gap remains: how to design nonequilibrium systems that achieve directed motion by biasing thermal fluctuations, without externally imposing any finite mean drift.

Herein, we introduce self-viscophoresis as a distinct class of autonomous motion mechanism inspired by biological strategies, operating through self-regulated fluctuation-rectification. Rather than directly generating propulsion, a particle consumes energy to create an asymmetric viscosity field in its surroundings, as illustrated in Fig. 1A. This field modifies the local dissipation landscape and thereby biases the statistics of the particle's thermal fluctuations, producing persistent directed motion.

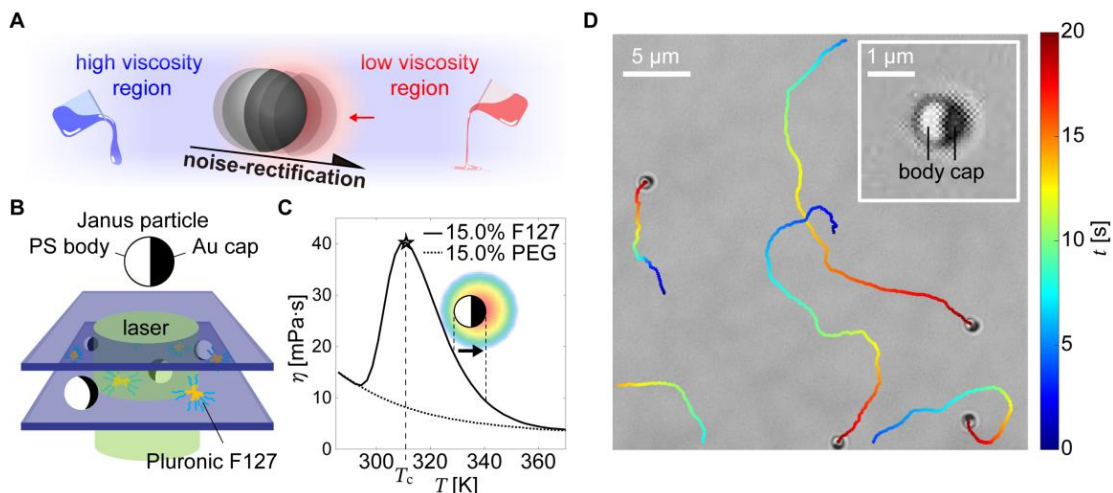

**Figure 1. Concept of the self-viscophoresis and experimental setup.** (**A**) Schematic illustration of self-viscophoresis. An asymmetric particle undergoes autonomous locomotion by generating a local viscosity gradient. This viscosity gradient rectifies the particle's stochastic dynamics, leading to a net displacement. (**B**) Experimental setup. Janus particles composed of a polystyrene (PS, white) core (diameter: 0.5–2 μm) and a 50 nm-thick gold (Au, black) cap are irradiated with a defocused 532 nm laser (spot diameter: 42 μm) in Pluronic F127 solution. (**C**) Temperature-dependent viscosity of 15.0% F127 (solid line) and 15.0% PEG (dashed line) solutions. In the F127 solution, a micellar network forms near room temperature, leading to a high viscosity. At elevated temperatures, dehydration-induced collapse of the micelles reduces the viscosity, which drops sharply above the critical temperature $T_C \approx 310$ K. Schematic of a heated Janus particle, where localized heating at the cap induces a temperature gradient, resulting in a strong viscosity contrast between the cap and body. (**D**) Representative trajectory under 360 mW laser, time color-coded. Inset: particle image (dark = cap, bright = body). Scale bars: main, 5 μm; inset, 1 μm.

In this paper, we experimentally demonstrate self-viscophoresis, where asymmetric colloidal particles in a thermoresponsive polymer solution generate viscosity gradients and undergo persistent directed motion upon uniform illumination. Systematic analysis revealed that the motion velocity is proportional to the diffusivity difference formed across the particle surface. To elucidate the mechanism underlying the observed motion, we developed a minimal Langevin model incorporating equilibrium thermal fluctuations coupled to a dynamically updating, nonequilibrium temperature–viscosity field. The model shows that this viscosity asymmetry biases otherwise isotropic thermal fluctuations into a net drift through anisotropic damping, reproducing the observed dynamics without invoking external forces or intrinsic self-propulsion. Finally, by tuning the temperature–viscosity characteristics of the polymer medium, we experimentally achieve reversible switching of both the direction and dimensionality of motion, thereby establishing a general and controllable framework for fluctuation rectification-based autonomous motion.

## RESULTS

### Laser-induced directed motion of Janus particles in a thermoresponsive medium

A classical laser-heating system for the Janus particle suspensions[16–19] was used in this study, as shown in Fig. 1B. The particles consisted of a micron/submicron-sized dielectric (polystyrene; PS) core half-coated with a thin metallic (gold; Au) cap. Upon laser irradiation, the cap induces localized heating owing to strong optical absorption. However, in aqueous solutions, such heating induces only minor changes in viscosity. To enhance thermal–viscous coupling, we introduced Pluronic F127[23–25], a temperature-responsive polymer exhibiting a lower critical solution temperature (LCST)-type sharp viscosity drop near its lower critical solution temperature $T_c \approx 310$ K, as indicated by the solid line in Fig. 1C.

The system base temperature was maintained at $T_{base} > T_c$, where the 15.0% F127 solution displayed monotonic temperature-dependent viscosity reduction. Under these conditions, the Janus particles, initially located at the bottom interface, exhibited persistent in-plane, directed motion, as shown in Fig. 1D. In contrast, under cooler conditions ($T_{base} \approx 300$ K), the particles moved out of the focal plane, likely due to radiation pressure[16,26] and vertical instabilities discussed later (see Section 1 of Supplemental Material). Figure 2A shows the representative trajectories of the persistent in-plane motion at $T_{base} = 318$ K $> T_c$ and various laser powers $I$ (see movie S1 of Supplemental Material). At $I = 0$ mW, the motion was purely random. At an intermediate laser intensity ($I = 420$ mW), cap-leading directed motion was observed. Interestingly, this directed motion vanished at $I = 540$ mW before reappearing with an opposite polarity at $I = 720$ mW, exhibiting a transition from cap-leading to body-leading motion. This complex, nonmonotonic response, including reentrant suppression and remarkable polarity reversal, deviates significantly from the classical self-propulsion types and suggests a new underlying mechanism. To quantify this dynamic behavior, we calculated the time-averaged squared displacement (TSD):

$$\overline{\delta^2(\tau; t_{total})} = \frac{1}{t_{total}-\tau} \int_0^{t_{total}-\tau}[\boldsymbol{R}(t+\tau) - \boldsymbol{R}(t)]^2 dt, \quad (1)$$

where $\tau$ is the lag time, $t_{total}$ is the total measurement time, and $\boldsymbol{R}(t)$ is the position of the center of the Janus particle as shown in Fig. 2B. The ensemble average over different particles for TSD $l(\tau) = \langle \overline{\delta^2(\tau; t_{total})} \rangle$ followed a power-law $l(\tau) \sim \tau^\alpha$, with $\alpha$ exhibiting a nonmonotonic dependence on laser power $I$: increasing, then dropping at $I = 540$ mW, then increasing again, indicating reentrant suppression as shown in Fig. 2C.

Because TSD does not capture motion polarity, that is, whether the motion is cap-leading or body-leading, we calculated the signed velocity $u = \boldsymbol{n} \cdot \boldsymbol{v}$, where $\boldsymbol{n}$ is the unit orientational vector pointing from the particle center toward the capped side determined via image analysis, and $\boldsymbol{v} = (\boldsymbol{R}(t+\tau) - \boldsymbol{R}(t))/\tau$ is the velocity vector over one frame interval with $\tau = 0.02$ s. This definition assigns a positive (negative) $u$ to cap-leading (body-leading) motion, enabling the direct quantification of polarity switching. Figure 2D shows a polarity reversal at $I \approx 540$ mW, marking a clear transition between two polarity regimes (solid line). This switching behavior is reversible under periodically modulated illumination and exhibits a rapid response (see Section 2 and movie 2 of Supplemental Material).

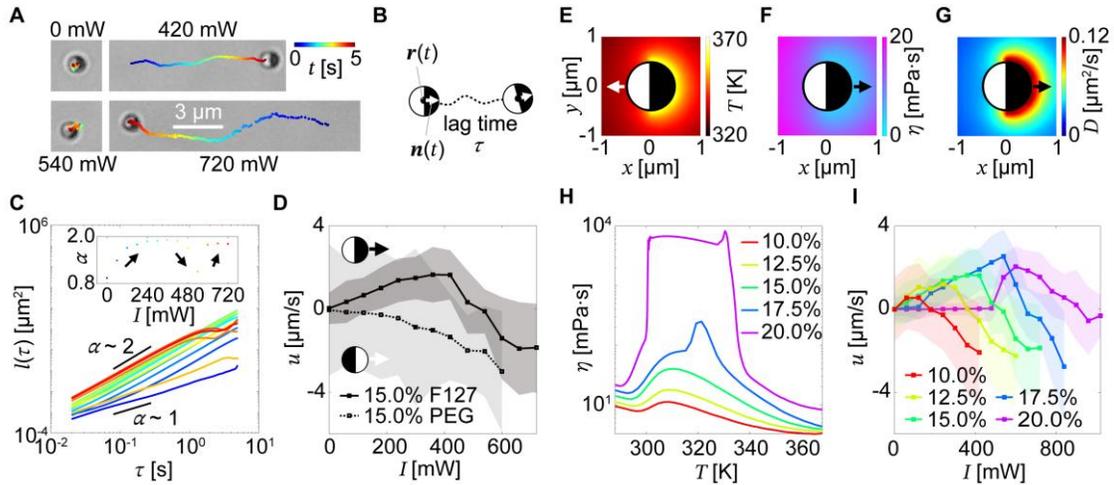

**Figure 2. Polarity switching of Janus particle motion and F127 concentration dependence.** (**A**) Representative trajectories in 15.0% F127 at $T_{base} = 318$ K $> T_c$, under various laser powers of $I = 0$, 420, 480, and 720 mW. Trajectories are color-coded by time. Scale bar: 3 μm. (**B**) Motion analysis schematic. $r(t)$ is the center-of-mass, and $n(t)$ is body-to-cap unit vector of the particle. (**C**) Ensemble-averaged TSDs $l(\tau)$ for $I=0$–720 mW. Inset: power-law exponent $\alpha$ vs. $I$. (**D**) Signed velocity $u$ vs. $I$ at $T_{base}=318$ K for 15.0% F127 (solid line) and 15.0% PEG (dashed line). Error bars: standard deviation. (**E** to **G**) COMSOL-calculated spatial profiles around a 1 μm-diameter Janus particle under $I=360$ mW and $T_{base}= 318$ K: (E) temperature distribution, (F) viscosity profile obtained by mapping via the temperature–viscosity relation in Fig. 1(C), and (G) diffusivity map derived from (E) and (F). (**H**) Temperature-dependent viscosity of F127 (10.0%–20.0%). (**I**) Signed velocity $u$ vs. $I$ for various F127 concentrations. Error bars: standard deviation.

To distinguish thermal- and viscosity-driven effects, we performed control experiments in a 15.0% polyethylene glycol (PEG) solution, which has similar molecular properties to F127 but lacks sharp viscosity transition, as indicated by the dashed line in Fig. 1C. Here, in PEG solution, the particles always exhibited body-leading motion ($u < 0$), and its negative value became stronger as $I$ increased, as shown by the dashed line in Fig. 2D. This behavior aligns with classical self-thermophoresis[16] driven by the Soret effect[27], potentially assisting the body-leading motion under a large temperature gradient $\nabla T$ and the positive Soret coefficient $S_\text{T}$ [16,28] of the particle. While this thermal mechanism might also explain the body-leading phase in F127, it fails to account for the observed cap-leading motion ($u > 0$), assuming a constant Soret coefficient.

One might argue, however, that a temperature-dependent reversal of the Soret coefficient in F127[29] could potentially drive such cap-leading motion. To examine this thermophoretic contribution, we introduced tracer nanoparticles to monitor the flow field around the Janus particle (see Section 3 and movie S3 of Supplemental Material). However, no directional solvent flow was detected for the cap-leading regime, thereby ruling out thermophoretic transport as the primary driver. Instead, the tracers enabled visualization of local diffusivity, revealing a marked increase in diffusivity around the cap side. This diffusivity enhancement originates from a local viscosity reduction induced by heating of the Janus cap, and such viscosity gradients have recently been shown to generate particle drift toward higher diffusivity region[30–34].

To estimate the spatial viscosity distribution $\eta$ and diffusivity distribution $D$, we used the Heat Transfer Modules of COMSOL Multiphysics (COMSOL Inc.; see Section 4 of Supplemental Material). First, we calculated the temperature distribution $T$, and subsequently, we derived the viscosity $\eta$ based on the experimentally measured temperature dependence of F127 shown in Fig. 1C. Finally, using the resulting spatial maps of $T$ and $\eta$, we calculated the local diffusivity $D$ via the Stokes–Einstein relation. At $I = 360$ mW, where cap-leading motion is experimentally observed, the estimated $T$ profile shown in Fig. 2E supported the body-leading motion through regular thermophoresis. Meanwhile, the viscosity profile shown in Fig. 2F exhibited a significant reduction near the cap, creating a diffusivity gradient $\nabla D$ shown in Fig. 2G. This gradient is expected to bias the particle drift toward higher diffusivity (lower viscosity) regions, thereby potentially giving rise to the observed cap-leading motion.

The viscosity-temperature relationship of the F127 solution depends on its concentration; higher concentrations increase the baseline viscosity, exhibiting greater viscosity contrast owing to solubility transitions, as shown in Fig. 2H. This causes a different mapping of $\eta$ and $D$, affecting the particle's motion. By analyzing the influence of these viscosity characteristics on particle motion, we can validate the mechanisms of cap-leading motion.

Across all concentrations, the signed velocity $u$ exhibited a similar trend with polarity reversal, but the transition point shifted to a higher laser intensity $I$ at higher concentrations, as shown in Fig. 2I. This reflected a balance between self-viscophoresis and self-thermophoresis: a steeper $\nabla \eta$ at higher F127 concentrations enhances cap-leading motion that overcomes thermophoretic body-leading drift even under strong illumination producing a large $\nabla T$. Interestingly, while larger peaks were observed generally at higher concentrations, the 20.0% solution was the exception, exhibiting a lower peak velocity. This was because, although a large viscosity contrast existed at high concentrations, the resulting increase in viscous drag suppressed the motion. These results suggest that cap-leading motion can be classified as self-viscophoresis, driven not simply by $\nabla \eta$ but potentially by diffusivity gradients $\nabla D$, which is linked to spatial variations in viscous damping.

### Scaling relations and universal self-viscophoretic response

To quantify the self-viscophoresis mechanism, we next examined how the velocity $u$ on the particle diameter $2a$. If self-viscophoresis is governed by $\nabla D$, $u$ should vary with particle diameter, as suggested by the Stokes–Einstein relation. We conducted experiments with Janus particles of diameter $2a = 0.50$–$1.50$ μm under weak illumination. In this weak intensity regime, the temperature gradient $\nabla T$ was minimal and the contribution from self-thermophoresis was negligible. In contrast, the viscosity transition of F127 persisted, enabling us to isolate the role of self-viscophoresis in driving particle motion.

In all the cases, $u$ increased monotonically with $I$, as shown in Fig. 3A, indicating that the self-thermophoretic contribution was negligible over these ranges. This effective range of $I$ varied with particle size because the heating efficiency scaled with $2a$. To account for this size effect, we normalized the laser intensity to $2aI$, which approximately equalized the temperature landscape across the different particle sizes, as shown in Fig. 3B. As initially expected, smaller particles exhibited higher

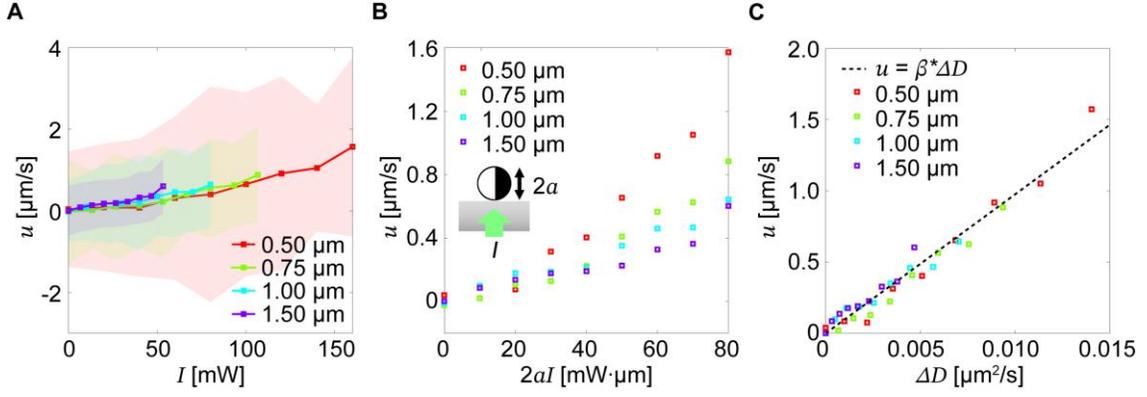

**Figure 3. Particle size dependence of self-viscophoresis. (A)** Signed velocity $u$ vs. laser power $I$ for Janus particles with diameters $2a$ = 0.50, 0.75, 1.00, and 1.50 μm under weak illumination. Error bars: standard deviation. **(B)** Replot of (A), with $I$ normalized by particle diameter ($2aI$). **(C)** Reconfigured data of (B), showing the signed velocity $u$ plotted against the diffusivity difference $\Delta D$ estimated from COMSOL-based diffusivity probing, yielding a normalized viscophoretic coefficient $\beta^*$ = 97.51 μm$^{-1}$.

velocities reflecting their ability to sense a steeper $\nabla D$ under a given $\nabla \eta$.

To quantify the relation between $u$ and $\nabla D$, we calculated $\nabla D \approx (D_{\text{cap}} - D_{\text{body}})/2a$, where $D_{\text{cap}}$ and $D_{\text{body}}$ are the local diffusivities at the immediate vicinity of cap and body pole, respectively, which both were estimated from simulated temperature probing (see Section 4 of Supplemental Material). Replotting $u$ from Fig. 3B against $\nabla D$ revealed linear trends, with slopes defining the viscophoretic coefficient $\beta$ (see Section 5 of Supplemental Material). Consistent with previous reports on passive viscophoresis[30–34], our result showed that larger particles exhibited larger $\beta$, reflecting their ability to monitor larger diffusivity differences across their surface. This suggests that this drift is more universally governed by the absolute difference $\Delta D \approx D_{\text{cap}} - D_{\text{body}}$, rather than the gradient $\nabla D$. Indeed, plotting $u$ against $\Delta D$ collapsed all data onto a single master curve, as shown in Fig. 3C, yielding a unified slope, which represents a normalized viscophoretic coefficient $\beta^*$ =97 ± 2 μm$^{-1}$, comparable to values reported in passive systems[33,34]. These findings support the validity and generality of the self-viscophoresis framework.

## Mechanism of self-viscophoretic motion

Having identified $\Delta D$ as the key quantity governing the drift, we examined how this diffusivity contrast, generated out of equilibrium, biases stochastic fluctuations to produce an effective net motion. To elucidate this mechanism, we performed two-dimensional stochastic simulations of a self-viscophoretic Janus particle using underdamped Langevin equations for translational and rotational motion.

$$m\frac{d^2\boldsymbol{R}(t)}{dt^2} = -f_{\text{sTP}}(\Delta T)\boldsymbol{n}(t) - \gamma^*\frac{d\boldsymbol{R}(t)}{dt} + \boldsymbol{\xi}(t), \quad (2a)$$

$$J\frac{d^2\theta(t)}{dt^2} = -\zeta^*\frac{d\theta(t)}{dt} + \nu(t), \quad (2b)$$

where $m$, $J$, $\boldsymbol{R}(t)$, and $\boldsymbol{n}(t)$ denote the particle's mass, moment of inertia, position, and the unit orientational vector $\boldsymbol{n}(t) = [\cos\theta, \sin\theta]$, which points from the particle center toward the capped side, respectively. The angle $\theta(t)$ represents the direction of the capped side measured from the positive $x$-axis. The translational and rotational thermal noises, $\boldsymbol{\xi}(t)$ and $\nu(t)$, are zero-mean white Gaussian noises with $\langle \xi_i(t)\xi_j(t')\rangle = 2\gamma^* k_B T^* \delta_{ij}\delta(t-t')$ and $\langle \nu(t)\nu(t')\rangle = 2\zeta^* k_B T^* \delta(t-t')$, respectively. Here, $k_B$ is the Boltzmann constant, and $T^*$ is the effective temperature. The first term on the right-hand side of Eq. (2a) represents the temperature-dependent self-propulsion force that drives the particle toward the body direction, which results from self-thermophoresis (sTP) driven by the temperature difference between the particle's cap and body, $\Delta T$, i.e., $f_{\text{sTP}} \propto \Delta T$. The second and third terms correspond to the diffusive motion, which consists of the damping term and the fluctuating term. The effective viscous drag coefficient $\gamma^*$ and effective rotational drag coefficient $\zeta^*$ are given by $\gamma^* = 6\pi a \eta^*$ and $\zeta^* = 8\pi a^3 \eta^*$, respectively, where $a$ is the particle radius. The effective viscosity $\eta^*$ is defined as the azimuthal average of the local viscosity field $\eta(\boldsymbol{r}, t)$ along the particle surface.

In our simulations, we considered various temperature-dependent viscosity profiles designed to qualitatively mimic the concentration dependence of F127 solutions, as shown in Fig. 4A. The local viscosity $\eta(r,t)$ was modeled as $\eta = \varepsilon \exp\left(-\frac{T(r,t)-T^{\text{offset}}}{\tau_t}\right) + \eta_\infty$, where $\varepsilon$, $T^{\text{offset}}$, $\tau_t$, and $\eta_\infty$ represent the scaling factor, temperature offset, temperature decay parameter, and baseline viscosity, respectively. The temperature field $T(r,t)$, which determined the local viscosity, was obtained using a continuum thermal simulation based on the heat diffusion equation:

$$\rho c_p \frac{\partial T(r,t)}{\partial t} = \kappa \nabla^2 T(r,t) + Q(r,t). \quad (3)$$

The heat source term $Q(r,t)$ consists of localized heating at the cap due to laser irradiation ($Q_{\text{heating}}$) and thermal dissipation in the surrounding substrate ($Q_{\text{dissipation}}$), and is defined as

$$Q(r,t) = Q_{\text{heating}} + Q_{\text{dissipation}}$$
$$= Q_c \chi(r) - \frac{1}{\tau_h}(T(r,t) - T_{\text{base}}), \quad (4)$$

where $\rho$, $c_p$, and $\kappa$ denote the density, heat capacity, and thermal conductivity of the particle and medium, respectively. $Q_c$ represents the heating strength corresponding to laser intensity, $\tau_h$ is the thermal relaxation time, and $T_{\text{base}}$ represents the base temperature at an infinite distance from the particle. The characteristic function $\chi(r)$ is introduced to distinguish the cap region of the Janus particle:

$$\chi(r) = \begin{cases} 1 & \text{in the cap region} \\ 0 & \text{otherwise} \end{cases}. \quad (5)$$

To isolate the effect of the self-induced viscosity gradient, we performed the simulations under the assumption that the density of the medium and the particle's orientation remain constant (i.e., $\rho = \text{const.}$, $\theta(t) = \text{const.}$) and scaled the thermal noise correlations using the effective viscous drag coefficient $\gamma^*$, such that $\langle \xi_i \xi_j \rangle \propto \gamma^*$. Notably, the thermal noise correlations were considered temperature-independent, which allowed us to emphasize the effect arising purely from self-induced viscosity asymmetry.

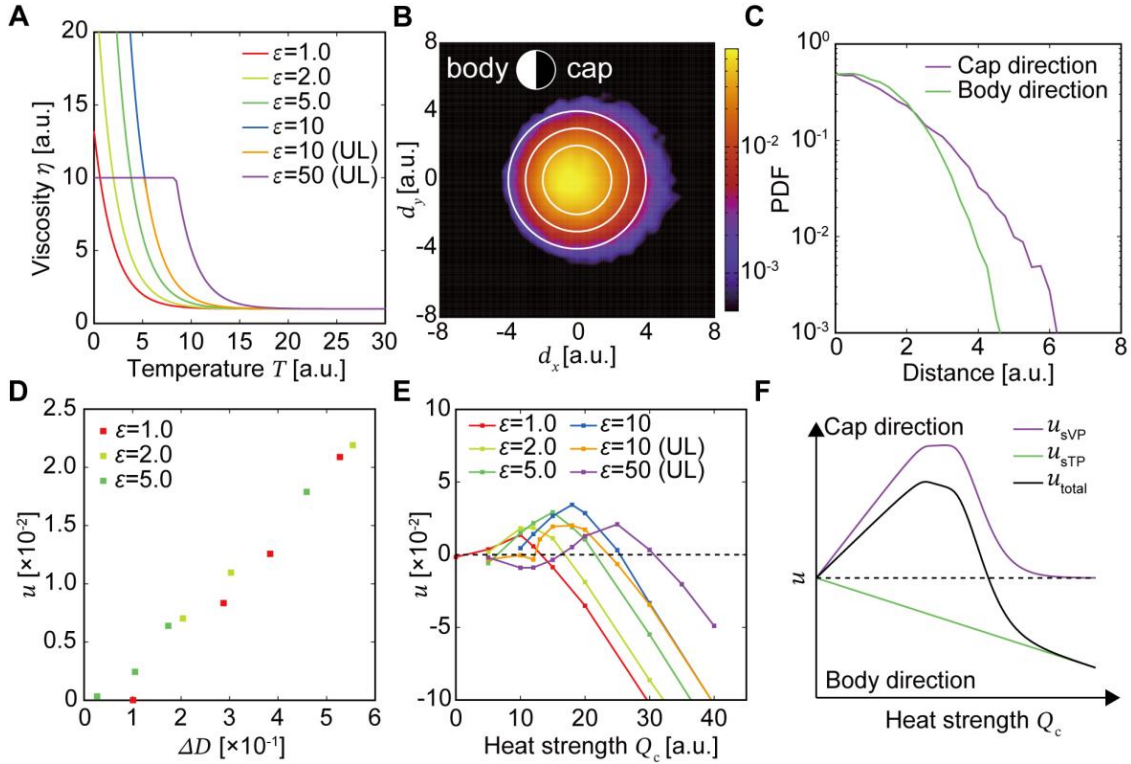

**Figure 4. Two-dimensional simulation of self-viscophoretic motion.** (**A**) Temperature-dependent viscosity profiles. Colored lines represent different scaling factors $\varepsilon$, with and without upper limits (UL) on viscosity $\eta_{UL} = 10$. (**B**) Probability density map of the particle's maximum displacement from its initial position. Positive $d_x$ corresponds to the cap direction and negative $d_x$ to the body direction of the particle. The white rings are added as visual guides to indicate distance from the center. (**C**) Probability distribution function of the maximum displacement. Color lines represent different directions. (**D**) Signed velocity $u$ vs. diffusivity difference $\Delta D$ under various viscosity profiles. (**E**) Signed velocity $u$ vs. heat strength $Q_c$ for different viscosity profiles. (**F**) Schematic of the dependence of both the self-viscophoresis and the self-thermophoresis on heat strength $Q_c$.

We first examined the particle behavior in a nonuniform viscosity field by conducting simulations without the self-thermophoresis effect (*i.e.*, $f_{\text{sTP}} = 0$). Figure 4B shows the probability density map of the maximum displacement of the particle calculated in the absence of thermal noise. The particle was initially placed in the steady-state viscosity field generated by the fixed particle and was given an initial random velocity sampled from a Gaussian distribution. Although both the initial particle velocity distribution and the azimuthal averaging used to define the effective viscosity were radially symmetric, the resulting particle displacement exhibited a directional bias toward the cap side, as shown in Figs. 4B and 4C. This bias arises from the finite thermal relaxation time $\tau_h$, which causes the viscosity field to follow the particle motion with a lag. This lag leads to asymmetric viscous damping even though the effective viscosity remains symmetric. Consequently, this lag-induced viscosity field enhances the persistence of inertial motion toward the lower-viscosity region near the cap side, whereas motion toward the body side is rapidly suppressed by stronger viscous damping. This effect is expected to be further amplified in the presence of thermal noise since the asymmetric damping continuously rectifies stochastic fluctuations into a net drift.

To clarify the effect of the viscosity gradient on the signed velocity $u$, we obtained $u$ from the simulation, including the thermal noise, as shown in Fig. 4D. In addition, the difference in the local diffusion coefficient between the cap and the body, $\Delta D$, was calculated using the Stokes–Einstein relation based on the steady-state viscosity distribution around a fixed particle. The signed velocity $u$ exhibited a linear dependence on $\Delta D$ for various viscosity profiles at low $Q_c$, as shown in Fig. 4D, which was consistent with the experimental results shown in Fig. 3C. This indicated that the cap-leading motion resulted from the asymmetry in the local diffusivity induced by the self-generated viscosity gradients.

The motility transition observed in the experiments shown in Fig. 2I was not reproduced in previous simulations that excluded the effect of self-thermophoresis. When the self-thermophoretic effect was included, i.e., $f_{\text{sTP}} \neq 0$, we observed a similar dependence of $u$ on the heating strength $Q_c$, as shown in Fig. 4E. Under all conditions, $u$ exhibited a positive peak as $Q_c$ increased and then decreased to negative values. The onset of this transition was progressively delayed as the scale factor $\varepsilon$ increased, reflecting the enhanced sensitivity of the viscosity to temperature. Furthermore, when the viscosity profile exhibited a plateau at low temperatures, $u$ remained nearly zero at low $Q_c$ and increased toward the peak beyond a certain threshold, as seen at $\varepsilon = 10(\text{UL})$ and $50(\text{UL})$ in Fig. 4E.

These simulation results suggest that the signed velocity $u$ can be interpreted as the sum of the self-viscophoresis velocity $u_{\text{sVP}}$ and self-thermophoresis velocity $u_{\text{sTP}}$, as shown in Fig. 4F. The self-viscophoresis velocity $u_{\text{sVP}}$ increases with increasing $Q_c$ in the low $Q_c$ regime, whereas it begins to decay at high $Q_c$ owing to a reduced viscosity gradient. This reduction is attributed to the heat generated at the cap reaching the body side through the surrounding medium and the particle body itself, thereby decreasing the viscosity contrast between the cap and the body (see Section 6 of Supplemental Material). In contrast, the self-thermophoresis velocity $u_{\text{sTP}}$ is proportional to the temperature difference between the cap and body. Consequently, the signed velocity $u$ exhibits a transition from the positive to the negative regime, which corresponds to a motion polarity transition.

### Diffusivity-contrast phase diagram of diverse 2D and 3D motility states

As discussed previously, the velocity $u$ of a Janus particle results from the competing effects of self-viscophoresis and self-thermophoresis, given by

$$u = \beta^* \Delta D - \frac{S_T D}{6a} \Delta T. \quad (6)$$

Here, the first term on the right-hand side represents self-viscophoresis, and the second term represents self-thermophoresis. The polarity of in-plane directed motion is determined by the sign of the lateral diffusivity difference $\Delta D_\parallel = \Delta D - \frac{S_T D}{6a\beta^*} \Delta T$, as illustrated in Fig. 5A. Meanwhile, the stability against vertical displacement is governed by the vertical diffusivity difference $\Delta D_\perp = D_{\text{Janus}} - D_{\text{base}}$ where $D_{\text{Janus}} = (D_{\text{cap}} + D_{\text{body}})/2$ is the averaged surface diffusivity, and $D_{\text{base}}$ is the bulk diffusivity, as illustrated in Fig. 5B. A positive $\Delta D_\perp$ stabilizes the particle at the interface, whereas a negative $\Delta D_\perp$ promotes bulk escape owing to radiation pressure induced by laser illumination from below.

Together, $\Delta D_\parallel$ and $\Delta D_\perp$ characterize four distinct motility regimes, tunable by laser power $I$ and base temperature $T_{\text{base}}$. When $T_{\text{base}} > T_c$, the viscosity of 15.0% F127 solution decreases monotonically with temperature, as shown in Fig. 1C, maintaining $\Delta D_\perp > 0$ and thus confining particles at the interface. In this regime, variations in $\Delta D_\parallel$ govern polarity switching via competition between self-viscophoresis and self-thermophoresis. In contrast, for $T_{\text{base}} < T_c$, viscosity increases with temperature, and $\Delta D_\perp$ can become negative, resulting in destabilized confinement and bulk migration. Here, even body-leading motion, which is typically attributed to self-thermophoresis at high $T_{\text{base}} > T_c$, can also emerge purely from viscous effect in this opposite trend, where a viscosity inversion leads to $D_{\text{cap}} < D_{\text{body}}$ and negative $\Delta D_\parallel$.

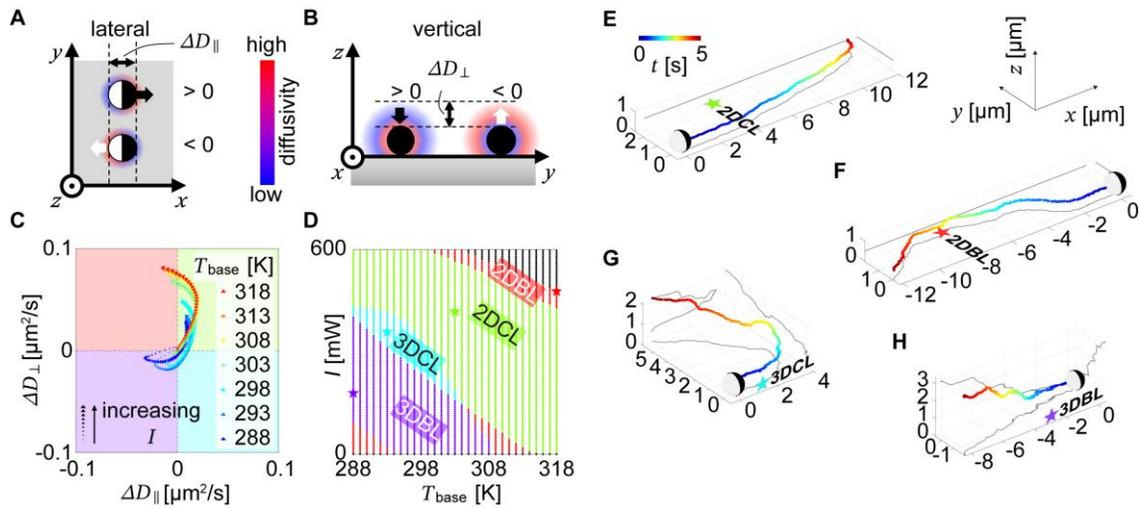

**Figure 5. Two-dimensional simulation of self-viscophoretic motion.** (**A**) Temperature-dependent viscosity profiles. Colored lines represent different scaling factors $\varepsilon$, with and without upper limits (UL) on viscosity $\eta_{UL} = 10$. (**B**) Probability density map of the particle's maximum displacement from its initial position. Positive $d_x$ corresponds to the cap direction and negative $d_x$ to the body direction of the particle. The white rings are added as visual guides to indicate distance from the center. (**C**) Probability distribution function of the maximum displacement. Color lines represent different directions. (**D**) Signed velocity $u$ vs. diffusivity difference $\Delta D$ under various viscosity profiles. (**E**) Signed velocity $u$ vs. heat strength $Q_c$ for different viscosity profiles. (**F**) Schematic of the dependence of both the self-viscophoresis and the self-thermophoresis on heat strength $Q_c$. **Diverse dynamics of Janus particles.** (**A**, **B**) Schematic of lateral (A) and vertical (B) diffusivity contrasts ($\Delta D_\parallel, \Delta D_\perp$); color indicates local diffusivity. (**C**) Calculated transition diagram in $\Delta D_\parallel$–$\Delta D_\perp$ space for a 1.00 μm-sized Janus particle in a 15.0% F127 solution, based on COMSOL diffusivity probing. Color and marker size encode base temperature $T_{\text{base}}$ and laser power $I$, respectively. (**D**) Motility-state diagram as a function of $T_{\text{base}}$ and $I$. The phase colors correspond to the quadrants in (C). Black region indicates $T_{\text{cap}} > 373$ K, where viscosity data is unavailable. (**E** to **H**) 3D trajectories of Janus particles exhibited various locomotion modes under following different ($T_{\text{base}}, I$) conditions I-IV: (E) 2D cap-leading (2DCL) at condition I (303 K, 420 mW), (F) 2D body-leading (2DBL) at condition II (318 K, 480 mW), (G) 3D cap-leading (3DCL) at condition III (293 K, 360 mW), (H) 3D body-leading (3DBL) at condition IV (288 K, 180 mW). 3D tracking was performed manually (see Section 8 of Supplemental Material).

Figure 5C shows motility-state trajectories in $\Delta D_\parallel - \Delta D_\perp$ space, obtained from COMSOL-based calculation for diffusivity probing. Each trajectory started at the origin and evolved with increasing $I$. At high $T_{\text{base}}$, the trajectories remained in the upper half-plane (i.e., $\Delta D_\perp > 0$), enabling polarity switching in-plane. As $T_{\text{base}}$ decreased, the paths extended into the lower quadrants ($\Delta D_\perp < 0$), indicating vertical instability.

To investigate the dependence of these motility states on the laser power $I$ and base temperature $T_{\text{base}}$, we numerically constructed a motility-state diagram, as shown in Fig. 5D. Each motility domain, defined by the sign combinations of $\Delta D_\parallel$ and $\Delta D_\perp$, was assigned a distinct color, corresponding to the four quadrants in Fig. 5C. Here, the transition between these four states was predicted to depend on both $I$ and $T_{\text{base}}$, even within the experimentally accessible range (0–600 mW and 288–318 K).

Experiments under four representative conditions, I–IV, indicated by the star symbols in Fig. 5D, validated these predictions (see movie S4 of Supplemental Material). Condition I exhibited in-plane cap-leading motion (2DCL), as shown in Fig. 5E; Condition II produced body-leading motion (2DBL), as shown in Fig. 5F. At a lower temperature (Condition III), the particle moved out of the plane into the bulk, displaying 3D cap-leading motion (3DCL), as shown in Fig. 5G. To capture this out-of-plane behavior, we performed real-time 3D tracking by manually adjusting the focus of the microscope (see Supplementary Section 7). Condition IV also resulted in 3D motion but in a body-leading configuration (3DBL), as shown in Fig. 5H. These results demonstrated that self-viscophoresis enables diverse motility states governed by dual imbalances in lateral and vertical diffusivity asymmetry, suggesting a controllable strategy for designing active matter systems.

## DISCUSSION

Our experiments demonstrate that an unusual directed motion overcoming conventional self-thermophoresis emerges under a sufficient viscosity contrast. Stochastic simulations combining the Langevin equation with the heat conduction equation qualitatively reproduced the observed motility trend. These results reveal that self-viscophoresis is a new autonomous motion mechanism that arises without the introduction of external fields or inherent self-propulsion terms. Hereafter, we discuss the uniqueness of self-viscophoresis from the following two viewpoints.

Self-viscophoresis belongs to the class of force-free self-propulsion[16–22], in that particles generate sustained directed motion without exerting any net force or torque on the surrounding medium. However, from a hydrodynamic perspective, self-viscophoresis differs fundamentally from classical force-free phoretic mechanisms such as self-thermophoresis[16,17], self-diffusiophoresis[18–21], and self-electrophoresis[22], which rely on momentum exchange with the surrounding solvent through interfacial slip flows. Instead, self-viscophoresis is more appropriately classified as a fluid-independent mechanism, often discussed in the context of "dry active matter[35]" or "autonomous passive phoresis," such as self-dielectrophoresis[14,15], in which directed motion arises from self-generated asymmetric landscapes that bias stochastic dynamics rather than from solvent-mediated hydrodynamic flows.

However, from an energetic view, this self-viscophoretic mechanism differs from conventional fluid-independent phoresis. While in conventional systems the particles consume energy directly to produce propulsive force, in our case, the input energy is used solely to induce an asymmetric viscosity field, which itself does not generate a direct propulsion. As supported by our findings, self-viscophoresis requires neither external forces nor intrinsic propulsion; instead, it arises purely from a statistical bias due to the rectification of thermal fluctuations, operating out of equilibrium.

Thus, self-viscophoresis can be regarded as a new mechanism that shares features with both major classes of conventional self-propulsion—force-free type and fluid-independent phoretic type—while fundamentally differing from them. Conceptually, it may also be classified as an "information swimmer[36–38]", a recently proposed idea describing a Brownian particle that achieves directional motion by rectifying thermal fluctuations through information processing. Specifically, this involves a cycle of sensing, memory, and feedback control: measuring its direction of motion, storing that information, and adjusting the viscous drag accordingly. While this concept has been mostly explored theoretically, several experimental demonstrations have been reported. However, all of them rely on external feedback implemented through computer-based tracking or control, which operates far from the particle[39,40]. Consequently, the particle merely remains a passive receiver rather than an active information-processing swimmer.

In contrast, self-viscophoresis could be considered the first artificial realization of a truly autonomous information swimmer. Importantly, all information processing in this system seems to be inherently embedded, occurring within a near-field, closed physical feedback loop, without any assistance from external systems. A possible scenario for the information-processing steps is as follows: the particle's position and internal orientation are recorded in the surrounding bulk memory (F127 solution) as local viscosity changes induced by laser heating. As the particle fluctuates, it experiences different viscous drag depending on its displacement, which can be effectively interpreted as the particle sensing its own direction of motion. Simultaneously, the particle's motion is subject to feedback via the viscous drag. In practice, however, these processes occur concurrently rather than sequentially, making it difficult to define a precise information-processing flow or to quantify energetic efficiency. For comparison with theoretical studies composed of multiple discrete short processes, pulsed-laser measurements may be preferable to continuous-wave lasers. Nevertheless, the present study suggests that an autonomous information swimmer works even when these instantaneous processes are effectively continuous. This interpretation is further supported by recent theoretical work by Yasuda *et al.*[38] based on the informational Onsager–Machlup principle, which demonstrates that information feedback can sustain a finite steady swimming velocity even in nonequilibrium steady states with effectively continuous measurement and feedback processes. Consequently, self-viscophoresis potentially offers a minimal experimental model for statistical thermodynamics and a universal framework applicable to autonomous biological cases, such as the bacterial motility that originally inspired this work.

Beyond its fundamental implications, the concept of self-viscophoresis also provides a practical design principle for functional microsystems. Because the direction and dimensionality of motion can be reversibly switched simply by tuning the laser power and the rheological response characteristic of the polymer, this mechanism enables dynamic control of active particles without any modification of their mechanical structure or chemical composition. Such fluctuation-based motion, governed by the interplay between stochastic behavior of particles and non-equilibrium viscosity fields, provides a robust and tunable strategy for constructing reconfigurable micromachines and adaptive soft materials. In this sense, self-viscophoresis connects stochastic thermodynamics with materials engineering by showing how thermal fluctuations can be systematically biased through self-generated nonequilibrium environments to serve as a design principle for active systems.

Extending this concept beyond a single particle, the viscosity field produced by one active unit can act as a communication channel to others in its vicinity, mediating information transfer through shared rheological environments. Such interactions could give rise to collective behaviors, including clustering[41–43] and highly ordered motions[44–50]. This mechanism may provide a tunable and robust route for designing autonomous micromachines that not only achieve autonomous motion but also perform complex group-level functions[51–54] through the coupling of stochastic dynamics to self-generated nonequilibrium rheological fields.

## CONCLUSION

We have introduced self-viscophoresis as a previously unidentified mechanism of autonomous motion, in which thermal fluctuations are rectified into directed motion through self-generated, nonequilibrium viscosity asymmetry. Experiments using asymmetric Janus particles demonstrate that the drift velocity is universally governed by the absolute diffusivity difference across the particle surface, $\Delta D$, establishing a unified scaling law for fluctuation-mediated transport. A minimal stochastic model coupling underdamped Langevin dynamics to a dynamically evolving viscosity field reproduces the observed behaviors, confirming that the interplay between thermal fluctuations and viscosity-induced asymmetric damping is sufficient to generate persistent motion under nonequilibrium conditions. Importantly, the sign of the in-plane diffusivity contrast $\Delta D_\parallel$ determines the polarity of horizontal motion, while the sign of the out-of-plane contrast $\Delta D_\perp$ selects whether the motion remains planar or extends into three dimensions; their combined tuning yields reversible polarity switching and 2D–3D transitions captured by a unified diffusivity-contrast phase diagram. This framework establishes self-viscophoresis as a versatile design principle for fluctuation-mediated microsystems and a general route to autonomous motion in complex fluid environments.

## MATERIALS AND METHODS

### Janus particle fabrication

Polystyrene-gold Janus particles were prepared using vapor deposition on monolayers of polystyrene microspheres. Aqueous suspensions of polystyrene spheres with diameters of 0.50, 0.75, 1.00, and 1.50 µm (Polybead® Polystyrene Microspheres, Polysciences Inc.) were deposited (20 µL) onto plasma-cleaned (plasma cleaner: Ion Bombarder PIB-10, Vacuum Device Inc.) glass coverslips (24 × 24 mm, Matsunami) and spin-coated at 2,000 rpm for 60 s (spin coater: Opticoat MS-A100, Mikasa Inc.) to form particle monolayers. A 50-nm-thick gold layer was then deposited on the monolayer surface using thermal evaporation (VE-2012, Vacuum Device Inc.) with a gold wire source (0.1 mm, Nilaco Corp.). The coated particles were detached from the substrate using ultrasonication in 5 mL of distilled water for 1 min and subsequently collected. To remove uncoated particles and detached gold fragments, we centrifuged 1 mL aliquots and discarded the supernatants. The Janus particles were then redispersed in 1 mL of either Pluronic F127 or PEG solution, prepared as described below, and stored at 4 °C for 1 day before use.

### Preparation of F127 and PEG solutions and rheological characterization

Aqueous solutions of Pluronic® F127 (Sigma-Aldrich, Mw~12,600 g/mol) were prepared by dissolving 4.0, 5.0, 6.0, 7.0, or 8.0 g of powder in 40 mL of distilled water to yield concentrations of 10.0, 12.5, 15.0, 17.5, and 20.0%, respectively. A 15.0% PEG solution was prepared by dissolving 6.0 g of polyethylene glycol (PEG 10,000, Merck-Sobuchardt, Mw = 9,000–11,250 g/mol) in 40 mL of distilled water. All solutions were refrigerated at 4 °C for 1 day. Viscosity measurements were performed using a rheometer (Discovery HR-2, TA Instruments) in the oscillatory shear mode to characterize the temperature-dependent viscoelastic response. Measurements were conducted during the heating process from 283 to 373 K at a heating rate of 2.0

K/min, with a strain amplitude of 100% and an angular frequency of 30.0 rad/s.

### Sample preparation, recording, and image analysis

Janus particles dispersed in polymer solutions were loaded into sealed sample chambers by sandwiching 20 μL of suspension between a glass slide and a coverslip (24 mm × 24 mm and 18 mm × 18 mm, respectively; Matsunami), separated by a 120 μm imaging spacer (SecureSeal™, Grace Bio-Labs, Sigma-Aldrich). The samples were observed using an inverted microscope (IX70, Olympus) equipped with a 100× oil-immersion objective, through which a 532 nm laser beam (Genesis MX532-5000MTM, Coherent) was introduced for defocused illumination. The base temperature $T_{\text{base}}$ was controlled using a stage heater (MI-IBC, OLYMPUS). Particle motion was recorded at 50 fps using a USB camera (WRAYCAM-NOA2000, Wraymer). Particle trajectories were extracted using the Particle Track and Analysis plugin in ImageJ, and custom MATLAB scripts were used to calculate the velocities and analyze the orientation of the gold cap from the bright-field images.

### Estimation of temperature distribution and particle dynamics simulation

The temperature field around the Janus particles under laser illumination was computed using the Heat Transfer Modules of COMSOL Multiphysics (COMSOL Inc.) to solve the steady-state heat conduction equation. Further details are provided in Supplementary Section 4.

### Stochastic simulation of particle dynamics

We conducted two-dimensional numerical simulations of a self-viscophoretic Janus particle coupled with a temperature-dependent viscosity field. The dynamics of the particles were modeled using the underdamped Langevin equation. The temperature field was obtained using a continuum thermal simulation based on a heat diffusion equation. Details of the model are described in the main text. Simulations were performed using the finite difference method for spatial derivatives and the explicit Euler method for time integration. All simulations shown in Fig. 4 were performed with a time step of $10^{-3}$. The system size was 256 × 256 and the spatial grid size was 1. The parameters used for Fig. 4 were $a = 10$, $T^{\text{offset}} = 5$, $\tau_t = 2$, $k_B T = 10$, $\kappa = 10$, $\tau_h = 10$, and $T_{\text{base}} = 0$. In the simulations, the effective viscous drag coefficient was set as $\gamma^* = a\eta^*$.

### ACKNOWLEDGMENT AND FUNDING

B. N. was supported by the Japan Society for the Promotion of Science (JSPS) KAKENHI (Grant Number JP24KJ0129) and the JSPS Core-to-Core Program "Advanced core-to-core network for the physics of self-organizing active matter" (JPJSCCA20230002). E. Y. was supported by JST PRESTO (Grant Number JPMJPR22EE). A. K. acknowledges the support from the Grant-in-Aid for Scientific Research (A) (Grant Number JP21H04434).

SUPPLEMENTAL MATERIAL

# Self-Viscophoresis: Autonomous Motion by Biasing Thermal Fluctuations via Self-Generated Viscosity Asymmetry

Bokusui Nakayama, Yusuke Takagi, Ryoya Hirose, Masatoshi Ichikawa, Marie Tani, Ibuki Kawamata, Eiji Yamamoto, and Akira Kakugo*

*Akira Kakugo
Email: kakugo.akira.8n@kyoto-u.ac.jp

**This PDF file includes:**

Supporting text from Section 1 to Section 7
Figures S1 to S7
Legends for Movies S1 to S4

**Other supporting materials for this manuscript include the following:**

Movies S1 to S4

Supporting Information Text

## 1. Experimental setup and vertical propulsion of Janus particles

We constructed a laser heating system for Janus particles, as illustrated in Fig. S1A. A continuous-wave laser beam (wavelength: 532 nm) was collimated using two lenses with focal lengths $f$ = 50 and 200 mm. A third lens ($f$ = 200 mm) was used to achieve defocused illumination. The resulting laser spot had a diameter of approximately 42 µm and was directed from below onto a sample containing Janus particles suspended in a Pluronic F127 aqueous solution.

At room temperature ($T_{base}$ = 300 K), the Janus particles exhibited vertical propulsion and departed from the focal plane upon laser illumination ($I$ = 180 mW) from below (Fig. S1B). Initially, we attributed this motion to the thermophoretic effects induced by local heating of the metallic cap. Supporting this hypothesis, symmetric light-absorbing particles such as 1.0 µm Dynabeads (iron oxide-embedded beads with high optical absorption) also exhibited similar upward motion (Fig. S1C).

However, we also found that non-absorbing polystyrene spheres (diameter 1.00 µm) displayed vertical propulsion under the same illumination conditions (Fig. S1D), suggesting that a non-thermal mechanism may be involved. To further examine the optical contribution, we tested silica particles with a refractive index closer to that of water and higher than that of polystyrene. The silica beads did not exhibit a vertical motion under identical conditions (Fig. S1E).

Type or paste text here. This should be additional explanatory text, such as: extended technical descriptions of results, full details of mathematical models, extended lists of acknowledgments, etc. It should not be additional discussion, analysis, interpretation, or critique.

These observations indicate that in addition to thermal effects, radiation pressure may contribute significantly to the vertical propulsion of particles. The sensitivity to the refractive index suggests that radiation pressure plays a role in driving the upward motion.

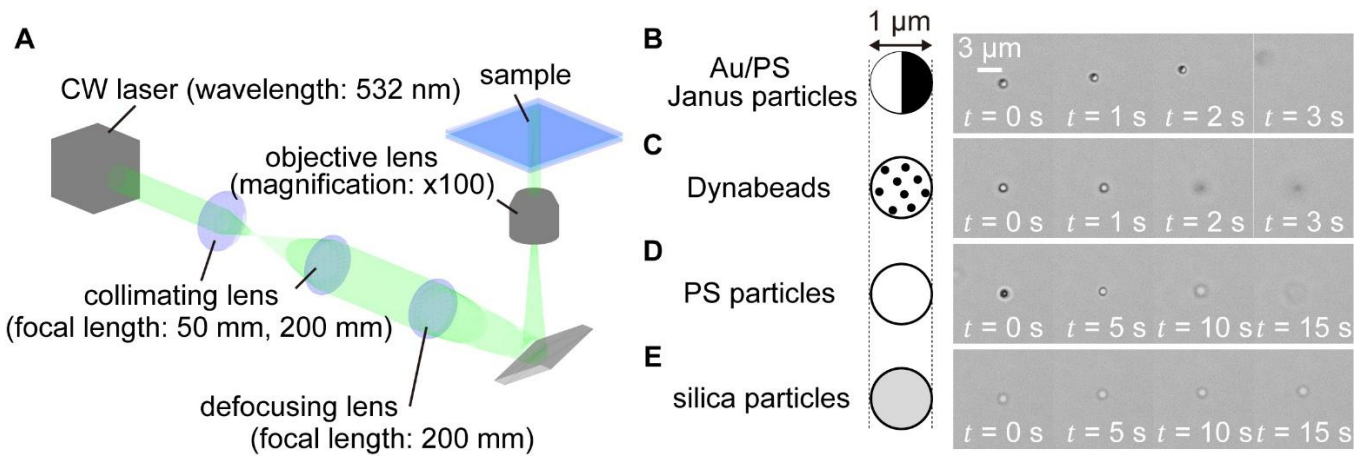

**Fig. S1. Optical setups and three-dimensional motion of particles with different material compositions under laser irradiation. (A)** Schematic of the experimental setup. A collimated and defocused laser beam creates a wide illumination spot. The following panels show representative snapshots of particles (diameter: 1.0 µm) under a base temperature $T_{base}$ = 300 K and laser power $I$ = 180 mW. **(B)** Polystyrene Janus particle with a gold cap exhibited cap-leading motion and departed from the substrate. **(C)** Dynabeads rapidly departed within a few seconds of illumination. **(D)** Polystyrene particle departed after approximately 10 s. **(E)** Silica particle didn't show detectable departure within 15 s of illumination.

## 2. Dynamic Polarity Switching of Janus Particle Motion under Periodically Modulated Laser Power

To demonstrate that the viscosity change process of F127 around the Janus particle is faster than the particle's motion, we investigated the dynamic response of the particle's locomotion to temporal variations in the laser power $I$. A Janus particle of 1 µm in diameter suspended in a 12.5% F127 solution was subjected to a periodically modulated laser power ranging from to 720 mW. Under these conditions, the particles exhibit periodic transitions between cap-leading and body-leading motions (Fig. S2A).

The time series of the applied laser power $I$ is shown in Fig. S2B, and the corresponding signed velocity $u$ of the particle is shown in Fig. S2C. At $I = 360$ mW. The particle displayed cap-forward motion with $u > 0$. As $I$ increased, the particle reversed its direction of motion and exhibited body-forward motion, corresponding to $u < 0$.

The time evolution of the signed path length $l(t) = \int_0^t u(t')dt'$ is presented in Fig. S2D. During the cap-forward motion, $l$ increased monotonically, whereas during the body-forward motion, $l$ decreased, reflecting a reversal in the net direction of motion.

These results indicate that Janus particles' motion responds rapidly and reversibly to changes in $I$. This suggests that the F127 viscosity change follows a fast, memoryless thermal relaxation process, likely owing to its underlying quick thermal conduction and molecular processes, including dehydration and micellar rearrangement.

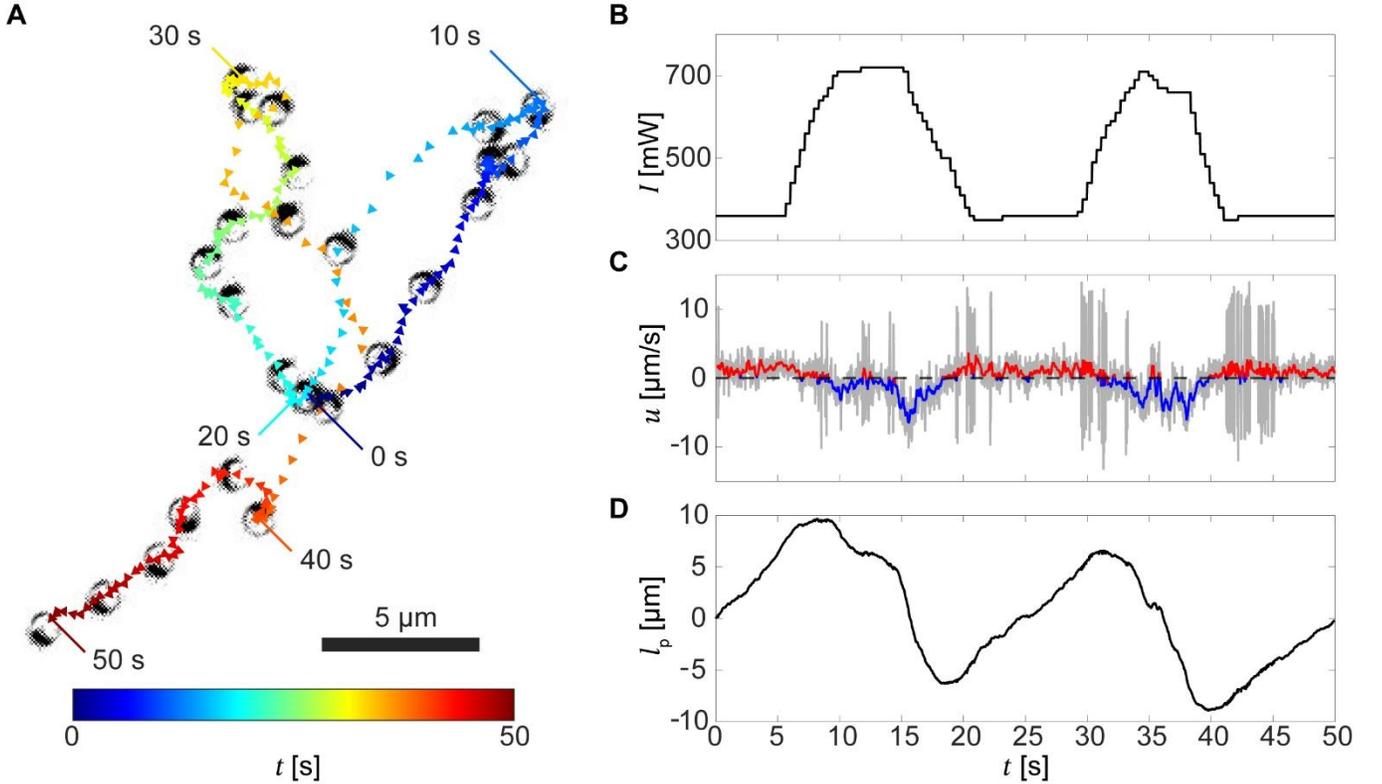

**Fig. S2. Dynamic switching of Janus particle motion polarity under periodically modulated laser power. (A)** Trajectory and overlaid snapshots of a Janus particle taken every 2 s under periodic laser power modulation in the 12.5% F127 solution. The trajectory is color-coded by time. The motion polarity of the particle switches approximately every 10 s. **(B)** Laser power modulation trace with a period of 20 s. **(C)** Time evolution of the signed velocity $u$. Gray dots indicate instantaneous values, and red (positive) and blue (negative) lines show a moving average over 0.2 s. The particle exhibits cap-leading motion ($u > 0$) at low $I$, and body-leading motion ($u < 0$) at low $I$. **(D)** Temporal evolution of the signed path length $l_\mathrm{p}(t)$, which varies periodically in response to changes in laser power $I$.

## 3. Visualization of Diffusivity Distribution around Janus Particles

In addition to the COMSOL-based estimations of the diffusivity distribution around the Janus particles, we experimentally visualized the local diffusivity profile using fluorescent nanodiamond (ND) tracers. By quantifying the thermal fluctuations of the NDs dispersed in the F127 solution, we indirectly mapped the spatial distribution of the local diffusion coefficient under laser illumination.

The laser played a dual role; it heated the Janus particles and excited the fluorescence of the NDs. To maximize the viscosity contrast around the Janus particles, we used a 20.0% F127 solution, exploiting the sharp sol–gel transition characteristic at this concentration (Fig. 2H).

A laser intensity of $I = 720$ mW was applied to the Janus particles. Although this power typically induces cap-forward propulsion of the particles (Fig. 2I), for precise local measurements, we selected Janus particles that adhered to the substrate and remained stationary during illumination.

Figure S3A shows representative trajectories of NDs undergoing Brownian motion near a laser-illuminated Janus particle. The NDs located far from the particle exhibited negligible motion owing to the high viscosity of the surrounding medium. In contrast, the NDs near the Janus particles exhibited pronounced fluctuations, reflecting a locally reduced viscosity.

To quantify the spatial variation in diffusivity relative to the Janus particles, we analyzed the image sequences from more than ten ($N > 10$) stationary Janus particles under continuous laser illumination. Each video frame sequence was cropped such that the Janus particle was centered, and then rotated such that the cap faced rightward in all cases. The resulting time-averaged images are presented in Fig. S3B. As expected, the mobile NDs appeared to be delocalized in the averaged image because of their fluctuations, whereas the stationary Janus particles remained centered with their cap consistently oriented to the right.

To capture the ND fluctuations, we calculated the time-integrated absolute difference between consecutive frames (frame-to-frame intensity difference) and generated the fluctuation maps shown in Fig. S3C. These maps revealed minimal changes in regions distant from the particle, whereas the fluctuation intensity increased toward the particle center, as shown in Fig. S3D.

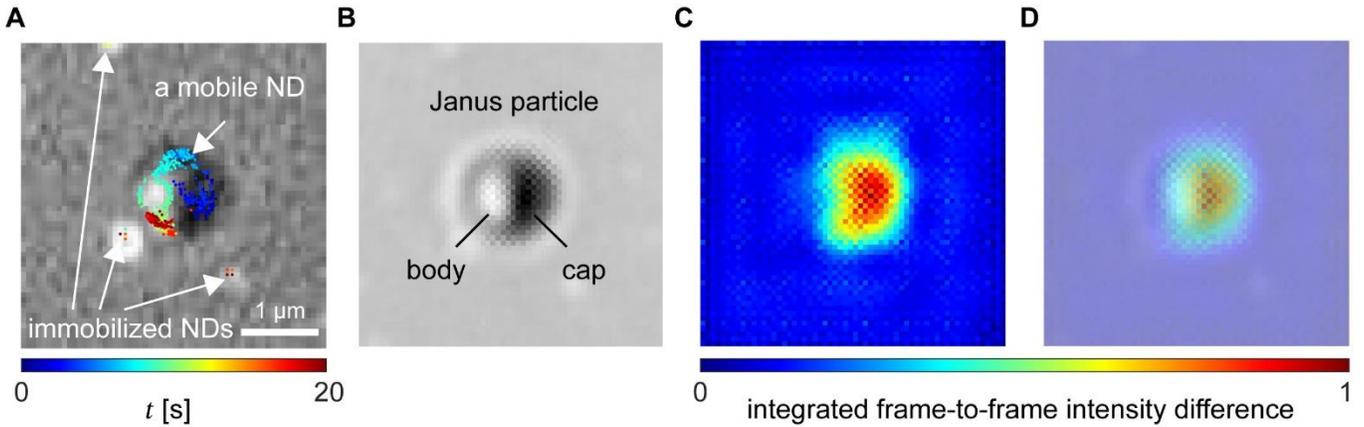

**Fig. S3. Direct visualization of local diffusivity around a Janus particle.** **(A)** Trajectories of nanodiamonds (NDs) near a Janus particle irradiated at $I = 720$ mW in 20.0% F127 solution. The trajectories are color-coded by time. NDs far from the Janus particle showed minimal fluctuations, while those near the particle exhibited enhanced Brownian motion. **(B)** Averaged image of Janus particles. Individual images of multiple Janus particles were aligned with the cap oriented to the right and averaged to obtain a representative configuration. **(C)** Diffusivity mapping around a Janus particle. The colormap represents the time-integrated intensity difference between consecutive frames, normalized across the field of view. **(D)** Overlay of panels (B) and (C), showing enhanced diffusivity near the capped side of the particle.

4. **COMSOL modeling of the temperature field calculation and the linear approximation method**

We used the Heat Transfer Module of COMSOL Multiphysics to estimate the steady-state temperature distribution around the Janus particles under laser illumination. The 3D simulation model is shown in Fig. S4A. The system consisted of three layers: the top and bottom layers represented glass substrates (material: $SiO_2$), each with dimensions of 200 µm × 200 µm × 200 µm (length × width × height). The sandwich between them was a water layer (material: $H_2O$) with dimensions of 200 µm × 200 µm × 120 µm, representing the sample chamber.

A Janus particle was placed at the center of the intermediate (water) layer, with its bottom surface in contact with the liquid–solid interface. The particle was modeled as a 1 µm diameter polystyrene sphere, half-coated with a 50 nm thick gold layer, and the cap was oriented horizontally. A top-hat laser beam with a wavelength of 532 nm, a spot diameter of 42 μm, and laser power $I$ was applied from below, centered at the bottom surface of the lower glass substrate.

Material properties (e.g., thermal conductivity, absorbance) were obtained from the default COMSOL materials library. A fixed temperature boundary condition was imposed on all external boundaries and set to the base temperature $T_\text{base}$ = 318 K. The steady-state temperature distribution was calculated, and temperature probes were placed at the gold cap and polystyrene pole of the Janus particle to extract the local temperatures $T_\text{cap}$ and $T_\text{body}$, respectively (Fig. S4B).

As the laser intensity $I$ increased, both $T_\text{cap}$ and $T_\text{body}$ increased linearly, with $T_\text{cap} > T_\text{body}$ consistently observed. The temperature differences relative to $T_\text{base}$, $\Delta T_\text{cap−base} = T_\text{cap} − T_\text{base}$ and $\Delta T_\text{body−base} = T_\text{body} − T_\text{base}$, were also found to scale linearly with I (Fig. S4C), with proportionality coefficients of $\xi_\text{c}$ = 0.116 K/mW and $\xi_\text{b}$ = 0.044 K/mW, respectively.

Assuming that the temperature increase caused by the gold cap is approximately independent of the absolute base temperature within the experimental range, we applied linear interpolation to estimate the local temperatures as a function of the laser intensity:

$$T_\text{cap} = T_\text{base} + \xi_\text{c} I, \quad (S1)$$
$$T_\text{body} = T_\text{base} + \xi_\text{b} I. \quad (S2)$$

These relations enabled us to estimate the local temperature gradient $\Delta T = T_\text{cap} − T_\text{body}$ for arbitrary laser intensities. Using the known temperature dependence of the viscosity of Pluronic F127, we further mapped the local viscosity distribution $\eta$ and corresponding diffusivity distribution $D$ around the Janus particle.

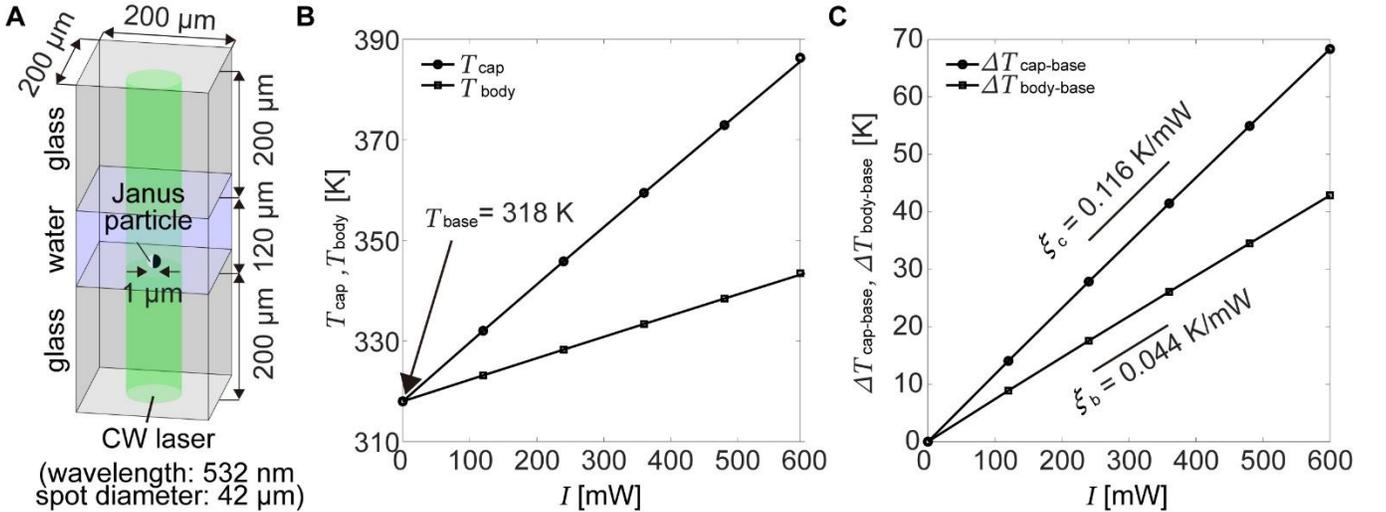

**Fig. S4. COMSOL model for estimating steady-state temperature profiles around a Janus particle. (A)** Geometry of the COMSOL model. The Janus particle is modeled in a configuration with the metallic cap facing horizontal direction. **(B)** Laser power $I$ dependence of the cap temperature $T_\text{cap}$ and body temperature $T_\text{body}$, probed at the poles of each hemisphere. The base temperature is set to $T_\text{base}$ = 318 K. **(C)** Temperature differences between the cap and base $\Delta T_\text{cap−base}$ and between the body and base $\Delta T_\text{body−base}$ as a function of $I$, derived from the data in panel b. These temperature differences scale linearly with $I$, enabling approximation of $T_\text{cap}$ and $T_\text{body}$ at arbitrary $I$ and $T_\text{base}$, via proportionality.

## 5. Relationship Between Janus Particle Size and Self-viscophoretic Coefficient

We re-plotted the data shown in Fig. 3B, i.e., the particle velocity $u$ of the Janus particles as a function of the normalized laser power $2aI$, with respect to the estimated diffusivity gradient $\nabla D$, based on COMSOL temperature simulations (Supplementary Section 4 and Fig. S4). Although larger particles are expected to exhibit higher thermal insulation due to the polystyrene core, resulting in an increased temperature difference $\Delta T$ between the cap and the body, this insulation effect was negligible at the particle size scale used in our experiments. Therefore, we assumed that at a fixed $2aI$, the cap and body temperatures were uniquely determined, regardless of the particle size.

For each particle diameter, we observed a proportional relationship between $u$ and $\nabla D$, with the proportionality constant obtained via linear fitting. This slope corresponded to the viscophoretic coefficient $\beta$, which was then plotted against particle diameter in Fig. S5B. The value of $\beta$ was observed to increase with particle size, indicating that larger particles exhibit stronger viscophoretic responses under identical thermal gradients.

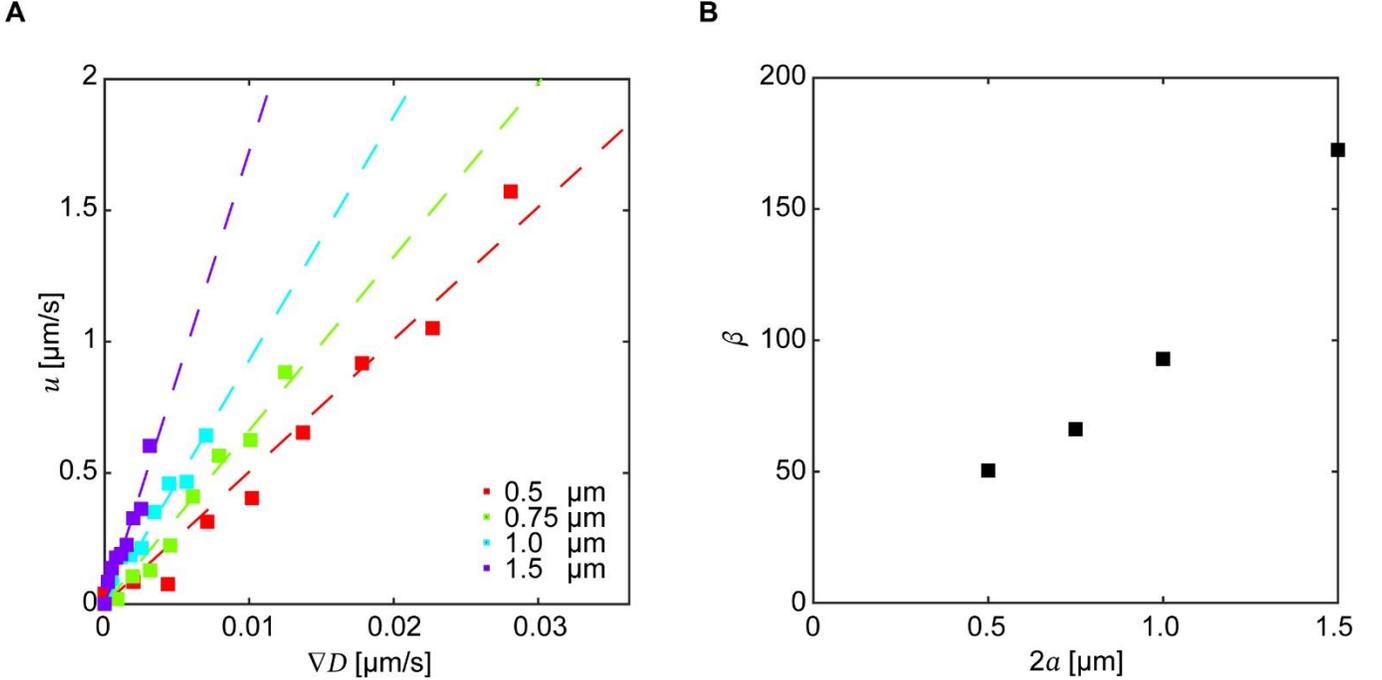

**Fig. S5. Relationship between Janus particle velocity $u$, diffusivity gradient $\nabla D$, and particle diameter. (A)** Velocity $u$ of Janus particles as a function of the diffusivity gradient $\nabla D$ for different particle diameters. The gradient $\nabla D$ was estimated based on the temperature field calculated using COMSOL simulations. Dashed lines represent linear fits to the data for each particle size. **(B)** Viscophoretic coefficient $\beta$, obtained from the slopes in (A), plotted as a function of particle diameter $2a$.

## 6. Simulated competition between self-viscophoresis and self-thermophoresis

Our experimental results suggest that Janus particles move under the competition between self-viscophoresis and self-thermophoresis. To examine this, we performed heat diffusion simulations to obtain the steady-state temperature and viscosity distributions around the particles. At low heat input (corresponding to low laser power experimentally), a heterogeneous viscosity distribution forms around the particle (Fig. S6A, top). The cap region is sufficiently heated, lowering viscosity, while parts of the particle body remain at high viscosity due to the insulating effect of the solvent and the particle core. By contrast, when the heat input increases, the low-viscosity region expands, surrounding the entire particle (Fig. S6A, bottom). This originates from the reduced temperature sensitivity of the F127 solution, which diminishes viscosity contrast (see the high-temperature region in Fig. 1C of the main text).

The time series of particle displacement along the x-direction (body to cap) is shown in Fig. S6B. At low heat input, the particle exhibits cap-leading motion, indicating dominance of self-viscophoresis driven by high viscosity contrast (Fig. S6B, top). At higher heat input, the reduced viscosity contrast combined with a stronger temperature gradient leads to body-leading motion driven by self-thermophoresis (Fig. S6B, bottom).

These results indicate that the observed polarity switching of Janus particle motion arises from the temperature-dependent properties of the polymer environment and the local temperature distribution. Proper polymer design and experimental conditions could thus enable more complex switching mechanisms and microscale particle control.

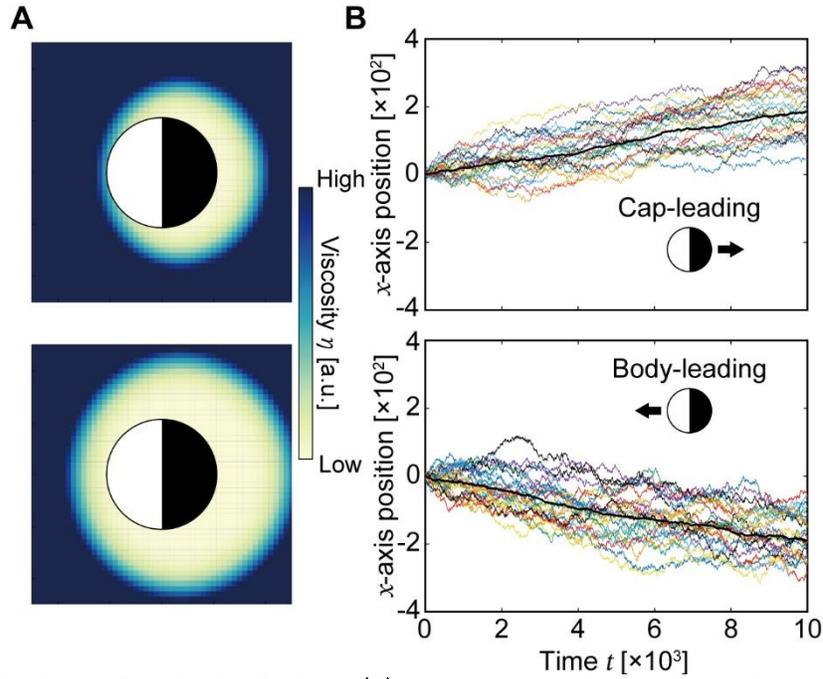

**Fig. S6. Motion polarity in stochastic simulations. (A)** Viscosity distribution around a particle under conditions that lead to cap-leading motion (top) and body-leading motion (bottom). The viscosity distribution was calculated when the position of particle was fixed. **(B)** Trajectories of particles in cap-leading motion (top) and body-leading motion (bottom). Positive $x$ corresponds to the cap direction, while negative $x$ corresponds to the body direction.

## 7. Three-Dimensional Tracking of Janus Particle Motion

We obtained the three-dimensional (3D) coordinates $(x, y, z)$ of Janus particles through a manual tracking method. To calibrate the axial displacement, we first determined that one full rotation of the microscope knob ($\Delta\varphi = 2\pi$) corresponded to a 13.5 μm axial displacement. To track knob angle $\varphi$ in real time, we attached a visual marker to the side of the knob (Fig. S7a). Here, the position of this marker was recorded and tracked using ImageJ, and its trajectory was fitted to a circular path using a custom MATLAB program to calculate $\varphi$ as a function of time.

By simultaneously recording both the particle image and the rotation of the microscope knob, we obtained time-resolved measurements of the particle's axial position $z_p$. However, owing to the manual operation of the knob, a certain delay and deviation inevitably occurred following the actual z-position of the particle. To quantify this error ($\delta z_p$), we analyzed the particle's image contrast and contour size as a function of defocus distance. Figure S7b shows typical defocused images of a Janus particle attached to the glass surface ($z_p = 0$ μm). When the focus was moved below the substrate ($\delta z_p < 0$), the particle image became dark and blurry with a reduced apparent contour diameter. Conversely, when focusing above the particle into the bulk ($\delta z_p > 0$), the image became bright and blurry, with an increased apparent contour diameter.

We manually measured the apparent contour diameter $R$ at different focus positions and correlated these values with the knob rotation-based $\delta z_p$ measurements. By fitting these data with a polynomial function, we obtained a calibration curve that allowed real-time estimation of $\delta z_p$ from the observed $R$. Figure S7c shows an example time sequence of $z_p$ and $z_p + \delta z_p$ determined using this method. Combining these $z$-position measurements with the x-y coordinates obtained from standard 2D tracking enabled full 3D tracking of the Janus particle trajectories (Fig. 5g–h).

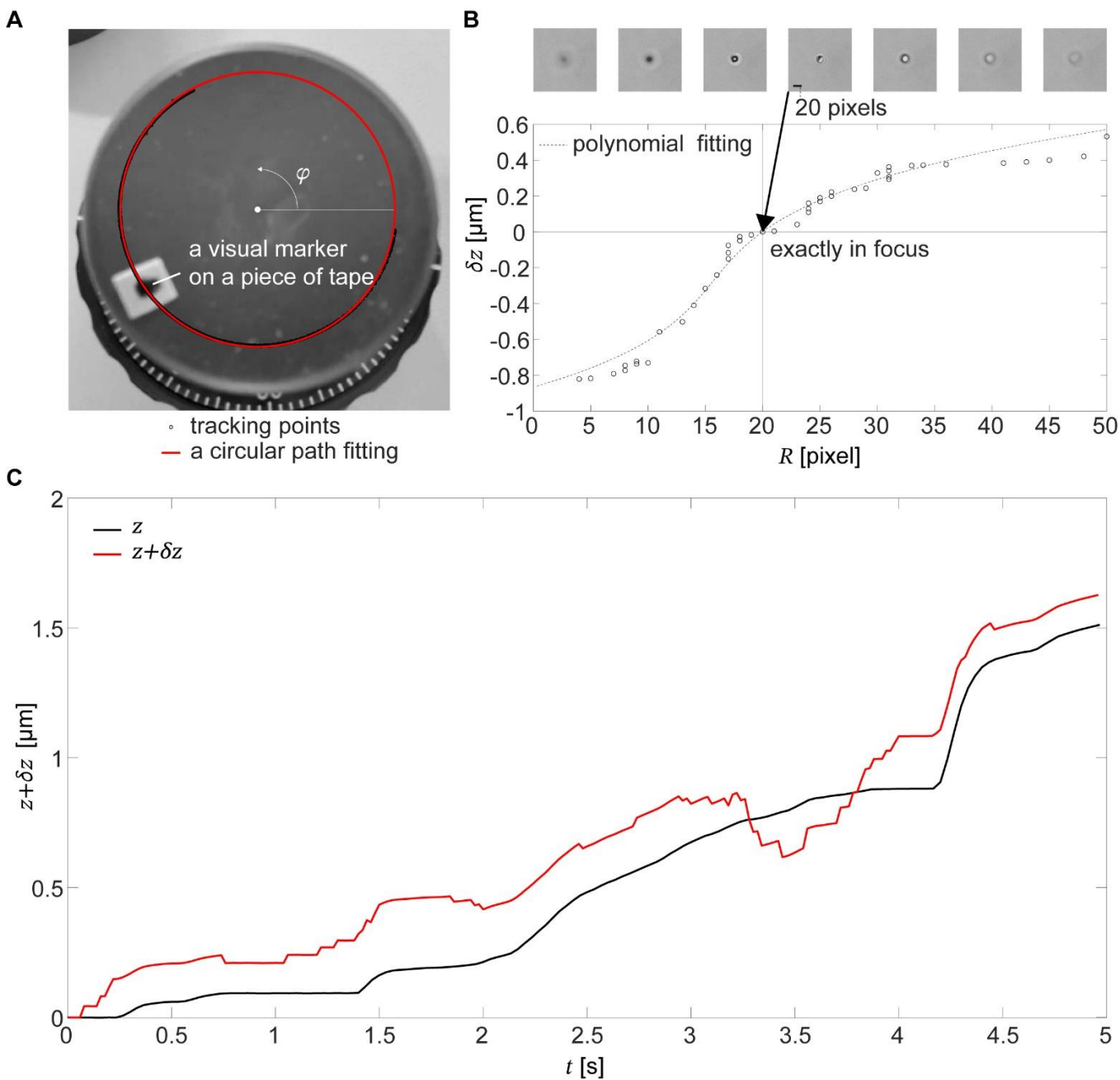

Fig. S7. 3D tracking method for Janus particle motion. (A) Calibration of microscope focus displacement. A marker was attached to the microscope focus adjustment knob, and its rotation angle $\varphi$ was recorded to determine the displacement of the focal plane. (B) Out-of-focus imaging of a Janus particle. A Janus particle immobilized on the substrate was imaged while the focal plane was shifted above and below the focal plane. The resulting defocused images were analyzed, extracting the dark ring. Its diameter $R$ was plotted as a function of the focal displacement $\Delta z$. (C) Representative time trace of the vertical $(z, z + \Delta z)$ position of a moving Janus particle. The z-coordinate estimated from the knob rotation (black) is compared with the corrected z-coordinate obtained from image-based defocus analysis (red).

## Supplementary Movies

**Movie S1 (separate file).** Polarity switching of Janus particle motion under laser illumination at different powers (Fig. 2A). The video shows the behavior of a Janus particle in 15.0% Pluronic F127 solution at a base temperature of $T_{base}$ = 318 K, recorded at 50 fps. Laser power $I$ was sequentially changed to 0, 420, 540, and 720 mW. Consequently, four distinct motility states were observed: random fluctuations, cap-leading propulsion, reentrant fluctuations, and body-leading propulsion, respectively. Playback speed: real-time (1×).

**Movie S2 (separate file).** Dynamic polarity switching of Janus particle motion under laser illumination with modulated power. A Janus particle in 12.5% Pluronic F127 solution at $T_{base}$ = 318 K was subjected to a laser whose power I was periodically modulated between 360 and 720 mW (approximately 20 s period). The particle exhibited cap-leading motion at 360 mW and body-leading motion at 720 mW. Videos were recorded at a rate of 50 fps. Playback speed: real-time (1×). Further details are provided in Supplementary Section 2.

**Movie S3 (separate file).** Visualization of local diffusivity around Janus particles. Fluorescent nanodiamond tracers (Sigma-Aldrich; diameter: 70 nm; excitation: 532 nm; emission: > 673 nm) were dispersed at stock concentration in 20.0% Pluronic F127 solution. The motion of the nanodiamonds near the surface-adhered Janus particles was recorded under laser illumination at $I$ = 720 mW and $T_{base}$ = 318 K. The videos were aligned such that the gold cap faced right. The tracers far from the Janus particles were immobilized in the gel-phase F127 medium, whereas those near the particles exhibited pronounced thermal fluctuations. Playback speed: real-time (1×). See Supplementary Section 3.

**Movie S4 (separate file).** Four motility modes of Janus particles. This movie presents representative examples of the four motility modes (2DCL, 2DBL, 3DCL, 3DBL) observed in 15.0% Pluronic F127 solution under various combinations of $T_{base}$ and I (corresponding to Fig. 5E to H). For the 3DCL and 3DBL states, manual 3D tracking was performed to capture particle trajectories. Playback speed: real-time (1×).